\documentclass{iopjournal}
\usepackage[numbers,square,sort&compress]{natbib}

\usepackage{bm}
\usepackage{float}
\usepackage{booktabs,tabularx,array}
\usepackage{graphicx,color}
\usepackage[dvipsnames]{xcolor}
\usepackage{enumerate}
\usepackage{harpoon}
\usepackage{amsmath,amssymb}
\usepackage{gensymb}
\usepackage{accents}
\usepackage[e]{esvect}
\usepackage[type={CC},modifier={by},version={4.0}]{doclicense}
\usepackage{footmisc}

\definecolor{ctcolor}{rgb}{0.8, 0.0, 0.2}

\begin{document}

\articletype{Paper} 

\title{Elastodynamics from Eulerian Poisson-bracket formalism: application to chiral odd solids}

\author{Cheng-Tai Lee$^{1,*}$\orcid{0000-0003-1916-039X} and Tomer Markovich$^{1,2}$\orcid{0000-0002-1887-0675}}

\affil{$^1$School of Mechanical Engineering, Tel Aviv University, Tel Aviv 69978, Israel}

\affil{$^2$Center for Physics and Chemistry of Living Systems, Tel Aviv University, Tel Aviv 69978, Israel}

\affil{$^*$Author to whom any correspondence should be addressed.}

\email{chengtailee@tauex.tau.ac.il}


\begin{abstract}
The Poisson-bracket (PB) formalism is widely used to derive dynamics of coarse-grained (CG) fields to capture large-scale physics, extending the role of PBs in classical particle mechanics to macroscopic field variables. It has been applied to study fluctuations in critical phenomena, hydrodynamics of liquid crystals, liquid crystal elastomers, tissues, and, more recently, the emergence of odd viscosity in suspension of spinning particles. The PB formalism can be formulated either in the Lagrangian framework, using the reference (undeformed) space, or in the Eulerian framework, using the real (deformed) space. Conventionally, the Lagrangian formulation is used for elastic solids, whereas the Eulerian formulation is for fluids. However, growing interest in Eulerian descriptions of solids has emerged for phenomena naturally defined in real space, such as viscoelastic responses, moving boundaries or interfaces, and field-induced structural changes in finite-sized particles. Here, we develop a systematic formulation for applying the Eulerian PB formalism to elastic systems whose potentials are typically written in Lagrangian space, and clarify its consistency with the Lagrangian PB formalism. We show that the Eulerian formulation generates additional nonlinearities absent in the Lagrangian counterpart. Such nonlinearities originate from CG volume changes under coordinate transformation, adding nonlinearities directly to the Hamiltonian, and from particle flow across neighboring CG volumes that result in nonlinear terms in the dynamics. They must be retained when nonlinear elastic effects are important. As an illustration, we study chiral active solids composed of finite-sized particles, where active torques drive internal particle rotations and thus generate geometric nonlinearities. These  nonlinearities give rise to the odd elastic modulus, which non-reciprocally couples two different shear modes in the stress-strain response. By recovering this modulus directly from the Eulerian PB formalism, we demonstrate its ability to capture emergent nonlinear elastic behavior in driven active solids, whose stresses are naturally measured in real space.
\end{abstract}

\section*{Introduction}

The Poisson-bracket (PB) formalism is a method for deriving dynamic equations of coarse-grained (CG) fields~\cite{chaikin1995,mazenko2006}. It extends the classical mechanics Poisson brackets, which gives the time evolution of particle-based observables~\cite{Goldstein_book} to mesoscopic field variables. This is done by applying the chain rule to the Hamiltonian derivatives and then obtaining PBs between CG fields from their microscopic definitions. Together with a coarse-grained Hamiltonian written in terms of these fields, the PBs generate the \textit{reactive} (\textit{non-dissipative}) contributions to the field dynamics, which obey time-reversal symmetry. In this CG description of dynamics, only a selected set of mesoscopic fields is retained to capture the large-scale physics. The microscopic degrees of freedom that are neglected give rise to dissipative effects, which must be obtained from another dynamical theory or from symmetry arguments.

The PB formalism offers several advantages. Once the mesoscopic fields are defined using microscopic particle quantities, it provides a systematic route to derive the corresponding CG dynamic equations. It also makes the physical origins of different terms in the field dynamics more transparent. Each contribution can be traced to specific PB couplings between the mesoscopic fields, whose form follows from the underlying particle variables. This provides a clear microscopic interpretation of the various terms in the dynamics. Moreover, the PB formalism can be applied not only to hydrodynamic variables, but also to non-hydrodynamic variables with finite (not system-size dependent) relaxation times. This allows one to rigorously eliminate these fast variables and obtain effective mesoscopic dynamics (see, e.g., Ref.~\cite{lubensky2005}). Finally, the PB formalism naturally generates nonlinear reactive terms, such as the streaming contribution in fluid dynamic equations. Combined with dissipative terms and  thermal noise, it provides a convenient framework for studying equilibrium dynamical correlation functions. This approach has been applied, for example, to study fluctuations in critical phenomena~\cite{kawasaki1970,dzyaloshinskii1980,mori1973,hohenberg1977}, where nonlinearities are central to the dynamics.

Due to these advantages and its generality, the PB formalism has been widely applied to diverse systems. Examples include flow alignment in liquid crystals~\cite{forster1974}, the coupling between strain rate and alignment in nematic polymers~\cite{kamien2000}, hydrodynamic equations of quasicrystals~\cite{lubensky1985}, liquid crystals~\cite{martin1972,stark2003,stark2005,kung2006}, liquid crystal elastomers~\cite{stenull2004}, and deformable particles with applications to tissue mechanics~\cite{hernandez2021,triguero-platero2023}. More recently, it has also been used to study the emergence of odd viscosity in chiral active fluids~\cite{markovich2021,markovich2024} and odd elasticity in chiral active solids~\cite{lee2026a}.
In these systems, the PB formalism is formulated in either Lagrangian or Eulerian coordinates. The \textit{Lagrangian} formulation uses the \textit{reference/undeformed} coordinate $\bm{r}$, whereas the \textit{Eulerian} formulation uses the \textit{real/deformed} coordinate $\bm{R}$. These coordinates are related through the displacement field $\bm{u}$ of the CG volume, $\bm{R}=\bm{r}+\bm{u}$. Conventionally, the Lagrangian PB formalism is used for elastic solids~\cite{stenull2004,lee2026a}, while the Eulerian PB formalism is used for fluids or fluid-like particles~\cite{forster1974,lubensky1985,kamien2000,stark2003,stark2005,hernandez2021,triguero-platero2023,markovich2021,markovich2024}. 

\begin{figure}[b!]
	\centering
	\includegraphics[width=8cm]{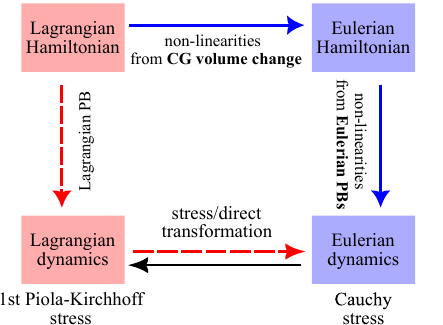}
\caption{Transformation between Eulerian and Lagrangian frameworks. Reference~\cite{lee2026a} applied Lagrangian PB formalism to a Lagrangian elastic potential and then used the stress transformation to get the Cauchy stress (red dashed arrow). Here we first find the Eulerian Hamiltonian and directly apply Eulerian PB  (blue solid arrow). Note that the Eulerian framework typically generates non-linearities. For example, a linear Lagrangian elasticity (i.e., with a quadratic potential in strain) results in a non-linear Eulerian elasticity due to the CG volume change between these two coordinates. Another source of non-linearities come from the Eulerian PBs, which create additional spatial gradient terms (Secs.~\ref{sec:field H} and \ref{sec:PBs} for details). Besides the stress transformation between the 1st PK stress and the Cauchy stress, in this work we also develop a direct transformation of the dynamic equations.}
	\label{fig:method}
\end{figure}

Recently, there has been growing interest in Eulerian formulations of elasticity. 
From a computational perspective, such formulations are advantageous in 
problems involving large deformations~\cite{Nave2012},
moving interfaces~\cite{valkov2015}, and fluid--structure interactions~\cite{liu2001,snoeijer2020,reinken2025,lee2026a}.
%
More fundamentally, an Eulerian description is natural in situations where the reference configuration is itself dynamic~\cite{ivanova2016}, such as in elastoplastic deformation~\cite{Rubin2019}, interfacial growth~\cite{naghibzadeh2021,Rubin2025}, and active materials, where elastic order coexists with rearrangements and flow~\cite{Henkes2024,Yang2024}. 
In these systems, and generally in viscoelastic materials (e.g.,~\cite{prost2015,Chen2023}), the distinction between solid-like and fluid-like behavior may evolve in time, making the    physical space (Eulerian) formulation particularly attractive. 
%
%
%
%
%
In this paper we apply the PB formalism to an elastic material in real (Eulerian) space, while clarifying the key differences between the Lagrangian and Eulerian PB formalisms. We further demonstrate their consistency in general, and specifically for chiral active solids, for which such a formulation may be particularly important in understanding its formation~\cite{tan2022}. 
%

The work is divided into a methodological part (Sec.~\ref{sec:formalism}) and an illustrative example (Sec.~\ref{sec:example}). In the first part, we develop the general Lagrangian and Eulerian PB formalisms for the field dynamics of finite-sized particles that can rotate and therefore carry angular momentum. We show that the Eulerian PB formalism generally produces higher-order terms absent from its  Lagrangian counterpart. These terms have two origins: the change of CG volume under the coordinate transformation, which gives rise to nonlinear terms in the Hamiltonian, and the additional spatial-derivative terms in Eulerian PBs, which account for particle flow from neighboring CG volumes and result in nonlinear terms in the dynamics. 

In linear elasticity these higher-order terms are neglected, and the two frameworks reduce to trivially equivalent forms. However, when nonlinear elastic effects become important,  Eulerian and Lagrangian dynamics becomes quite different.  
%
%
This can occur in materials under large deformation, which can be realized either through externally-induced strain or particle rotations, or by internal \textit{active} forces~\cite{marchetti2013,julicher2018,liebchen2022}. 
%
To demonstrate the consistency between the two frameworks, we complete the procedural loop connecting them, see Fig.~\ref{fig:method}. This includes a direct transformation between the dynamic equations derived in the Lagrangian and Eulerian frameworks. Using this transformation, we also recover the standard relation between the Cauchy stress and the first Piola-Kirchhoff stress tensors.
For simplicity and clarity, we present the main formulation in two-dimensional (2D) systems. The extension to three dimensions is discussed along the way and also in  Appendices~\ref{app:canonical conjugate pair} and \ref{app:Eulerian PBs}.

In the second part, we use 2D disordered chiral active solids~\cite{lee2026a} to illustrate the nonlinear effects generated by the Eulerian PB formalism and to demonstrate its consistency with the Lagrangian PB approach. In this system, chirality is introduced through internal active torques that drive particle rotations (see Sec.~\ref{sec:example} for details). These rotations induce \textit{geometric nonlinearities} in the elastic potential, coupling particle center-of-mass (CM) displacements to internal rotations, which we describe using micropolar (Cosserat) elasticity~\cite{eringen1966,eringen1999,eremeyev2013,lee2026a}.
These geometric nonlinearities are essential for the emergence of an odd elastic modulus in real space~\cite{lee2026a}, which couples  non-reciprocally two different shear modes in the stress-strain response~\cite{scheibner2020,braverman2021,fossati2024,fruchart2023,banerjee2025,lee2026a}. In 2D, a simple shear strain induces a pure shear stress, whereas a pure shear strain induces a \textit{negative} simple shear stress. Such non-reciprocal coupling leads to unusual mechanical responses, including tilting under uniaxial compression~\cite{scheibner2020,fruchart2023,lee2026a}, growing bulk wave modes~\cite{scheibner2020,lee2026a}, propagating bulk waves even in the overdamped regime~\cite{scheibner2020,lee2026a}, and unidirectional surface modes~\cite{abanov2018,souslov2019,soni2019,scheibner2020a,fossati2024,veenstra2025a,gao2022,caprini2025a,lee2026}.

\section{General formalism}\label{sec:formalism}  

\subsection{Coarse-grained fields}\label{sec:CG fields}

To apply the PB formalisms, one has to first define the macroscopic CG fields from microscopic particle quantities. These CG fields will then be used to formulate the system's Hamiltonian at large scales. To understand the differences and transformation between the Lagrangian and Eulerian frameworks, it is necessary to categorize these CG fields into density type and particle-averaged type. 
The density-type fields are defined as the sum of particle quantities within the CG volume via:
\begin{align}
\bm\Psi^\circ (\bm{r})
&\equiv 
\sum_{\alpha \in \Delta V^\circ} 
{\bm \psi}^\alpha \delta(\bm{r}-\bm{r}^\alpha)
\, , \label{eq:Lagran density fields}
\\
\bm\Psi(\bm{R}) 
&\equiv 
\sum_{\alpha \in \Delta V} 
\bm\psi^\alpha\delta(\bm{R}-\bm{R}^\alpha)    
\,  , \label{eq:Eulerian density fields}    
\end{align}
where $\delta(\cdots)$ is the Dirac delta function and $\alpha$ is the particle index. For the Lagrangian (Eulerian) framework, we use the reference (current) space before (after) deformation $\bm{r}$ ($\bm{R}$) and the CG volume $\Delta V^\circ$ ($\Delta V$) for the CG field $\bm\Psi^\circ(\bm{r})$ ($\bm\Psi(\bm{R})$). 

\begin{figure}[t]
	\centering
	\includegraphics[width=7.5
cm]{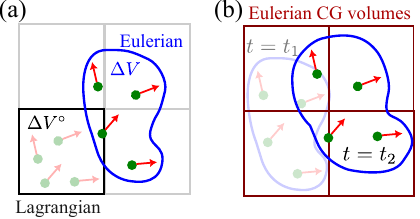}
\caption{(a) Schematic of the change in the CG volume from Lagrangian (reference) to Eulerian (deformed) space. Consider tracing the same pack of particles within the Lagrangian CG volume $\Delta V^\circ$ (black solid square). In Lagrangian space, these particles remain confined within this CG volume, and crossing between neighboring CG volumes is not allowed. {{The particles have other properties such as their linear momenta (in the schematic red arrows indicate their velocities).}} When the Lagrangian CG volume is mapped into Eulerian space, it occupies a different volume, $\Delta V=J\Delta V^\circ$ {{(blue enclosure)}}, as a result of the deformation, where $J\equiv\det(\bm{F})$. Since the same pack of particles is traced, their total physical quantities, such as mass and linear momentum, remain identical within the two corresponding volumes, $\Delta V^\circ$ and $\Delta V$. {Note that the deformed Eulerian volume is shown only to illustrate the relation between the Lagrangian and Eulerian fields and does not represent the actual Eulerian CG volume.} (b) In Eulerian space, particles can cross between neighboring CG volumes over time. For example, a particle located in the top-left CG volume (top-left brown square) at time $t=t_1$ crosses into the top-right CG volume at time $t=t_2$. This particle crossing gives rise to the spatial-gradient terms coupled to the linear-momentum dynamics $\dot{\bm g}_c$ in the Eulerian PBs {{(the most common example for this is the advection term)}}.
}
	\label{fig:field wrt volume change}
\end{figure}

Importantly, the superscript $\circ$ in $\bm\Psi^\circ$ is reserved for the \textit{Lagrangian} density-type fields. They are different from their physical, \textit{Eulerian} counterparts because deformation alters the CG volume (Fig.~\ref{fig:field wrt volume change}), such that $\Delta V = J \Delta V^\circ$. Here $J$ is the Jacobian of the coordinate change from $\bm{R}$ to $\bm{r}$, i.e., $J\equiv \det (\bm{F})$ with the deformation gradient tensor $F_{ij}\equiv \nabla^\circ_j R_i$ and $\nabla^\circ\equiv\partial/\partial r_j $. To find the transformation between the Lagrangian and Eulerian density-type fields, we trace the same pack of particles in both spaces (Fig.~\ref{fig:field wrt volume change}(a)). The two corresponding volumes therefore contain the same total physical quantities from these particles, such as their number, mass and linear and angular momenta. This then leads to $\bm\Psi^\circ \Delta V^\circ = \bm\Psi \Delta V$ when expressed in terms of the CG fields, and hence, 
\begin{equation} \label{eq:density field transformation}
\bm\Psi^\circ(\bm{r})
= J\ \bm\Psi (\bm{R})
\, .    
\end{equation}
The density-type fields include the densities of particle number ($n^\circ$, $n$), particle mass ($\rho^\circ$, $\rho$), moment-of-inertia ($I^\circ$, $I$), center-of-mass (CM) momentum ($\bm{g}^\circ$, $\bm{g}^c$), and angular momentum  ($\ell^\circ$, $\ell)$.  

In contrast, the particle-averaged fields are free of such volume-change effect and are defined via:
\begin{align}\label{eq:particle-averaged fields}
\bm\Phi (\bm{r})
&\equiv 
\sum_{\alpha \in \Delta V^\circ} 
\frac{{\bm \phi}^\alpha \delta(\bm{r}-\bm{r}^\alpha)}{n^\circ(\bm{r})}
\notag \\ 
=
\bm\Phi(\bm{R}) 
&\equiv 
\sum_{\alpha \in \Delta V} \frac{\bm\phi^\alpha\delta(\bm{R}-\bm{R}^\alpha)}{n(\bm{R})}    
\, ,     
\end{align}
with $n^\circ(\bm{r})$ and $n(\bm{R})$ being the particle number density before and after deformation, respectively. The particle-averaged fields include the CM displacement $\bm{u}$ and internal rotation $\theta$, which are typically used to write the elastic potential~\cite{LLelastity,eringen1966,eremeyev2013}. Note that particle-averaged fields can be defined using other ratios of density fields. For example, the CM velocity $\bm{v}^c$ is the ratio between the linear momentum density and the particle mass density $\bm{v}^c \equiv \bm{g}^\circ/\rho^\circ = \bm{g}^c/\rho$, while the angular velocity $\Omega$ is the ratio between angular momentum density and the moment of inertia density $\Omega \equiv \ell^\circ/I^\circ = \ell/I$. 




Note that the fields as defined in this subsection are discontinuous. The CG procedure, which is detailed elsewhere~\cite{markovich2024}, makes them smooth, but the relation to the microscopic particle quantities is well defined. This will be used in Sec.~\ref{sec:PBs} below.

\subsection{Hamiltonian} \label{sec:field H}


Using the CG fields, we formulate the \textit{Lagrangian} Hamiltonian 
$H$ at the continuum scale, which is composed of the kinetic energy density and the total potential density $V^\circ_\text{tot}$:
\begin{align}\label{eq:H}
H=\int d\bm{r} \
\bigg(
\frac{\left({\bm g}^{\circ}\right)^2}{2\rho^\circ} 
+ \frac{\left(\ell^{\circ}\right)^2}{2 I^\circ}
+ V^\circ_\text{tot} 
\bigg)
\, ,
\end{align}
where the first term is the CM kinetic energy (${\bm g}^\circ$ being the CM momentum density), and the second term is the rotational energy~\cite{markovich2024,Goldstein_book} arising from a \textit{finite} particle size ($\ell^\circ$ is the angular momentum density). This rotational energy is absent for point-like particles, as assumed in conventional elasticity theory, where particle motion is described only by the CM displacement $\bm{u}$~\cite{LLelastity}. The total potential density $V^\circ_\text{tot}$ includes an elastic potential $V^\circ$ as well as other types of potentials, such as those induced by external driving fields or internal active sources (see also Sec.~\ref{sec:potential}).

The rotational contribution in Eq.~\eqref{eq:H} is needed to derive the dynamics of rotation-related fields, which are usually non-hydrodynamic, but are pivotal when driven by activity~\cite{lee2026a}. For example, in nematic liquid crystals (LCs), in the hydrodynamic limit, $\ell^\circ=-I^\circ(dn/dt)$, where $n$ is the nematic director~\cite{stark2003,stark2005}, such that the angular momentum dynamics are determined by the director. 
In micropolar (or Cosserat) elastic media~\cite{eringen1966,eringen1999,eremeyev2013}, particles may also undergo internal rotations, described by the field $\theta$, and therefore carry angular momentum. In this case, the internal rotation is non-hydrodynamic and usually relaxes fast to a local mechanical equilibrium orientation~\cite{chaikin1995}, see also the example studied below in Sec.~\ref{sec:example}. 
Note that Eq.~\eqref{eq:H} can be generalized to 3D by using the tensor form of the moment-of-inertia density and the vector form of the angular momentum density. The rotational energy then takes the form $\ell^\circ_i (I^\circ)^{-1}_{ij}\ell^\circ_j/2$, where $(\bm{I}^\circ)^{-1}$ is the inverse of the moment-of-inertia density tensor $\bm{I}^\circ$. 

In order to write the Lagrangian Hamiltonian $H$ in the Eulerian framework, we change coordinates from $\bm{r}$ to $\bm{R}$ in Eq.~\eqref{eq:H} and use Eq.~\eqref{eq:density field transformation} to replace the Lagrangian density fields $\bm\Psi^\circ(\bm{r})$ by their Eulerian counterparts $\bm\Psi(\bm{R})$:
\begin{align}\label{eq:H Eulerian}
H
=
\int d\bm{R}\ \  
J^{-1} 
\bigg(
\frac{(g^{\circ})^2}{2\rho^\circ} 
+ \frac{(\ell^{\circ})^2}{2 I^\circ}
+ V^\circ_\text{tot} 
\bigg)
=
\int d\bm{R} \
\bigg(
\frac{(g^{c})^2}{2\rho} 
+ \frac{\ell^{2}}{2 I}  
+ V_\text{tot} 
\bigg)
\, .
\end{align}
$J^{-1}\equiv  \det(\bm{F}^{-1})$ is the inverse Jacobian factor, with the inverse deformation gradient  $F_{ij}^{-1} \equiv \nabla_j r_i$ and $\nabla_j \equiv \partial/ \partial R_j$. The $J^{-1}$ factor due to the coordinate change cancels the $J$ factor from the field transformation (Eq.~\eqref{eq:density field transformation}).

Importantly, from Eq.~\eqref{eq:H Eulerian}, we have $V_\text{tot}=J^{-1}V^\circ_\text{tot}$, such that the Eulerian potential energy density contains additional terms of higher order in $\nabla_j u_i$ (compared to $V^\circ_\text{tot}$).
%
In \textit{linear} elasticity, namely for small deformations and small internal rotations ($|\nabla_j u_i|$ and $|\theta|\ll 1$, respectively) these terms are negligible. Then, the Lagrangian and Eulerian potential densities, $V_\text{tot}^\circ$ and $V_\text{tot}$, have the same form.
However, when nonlinearities become important, the two potential energy densities generally differ, as illustrated later by the example of systems driven by active torques in Sec.~\ref{sec:example}.


\subsection{Poisson-bracket formalisms}\label{sec:PBs}


With the definitions of the CG fields and the expression for the CG Hamiltonian, the reactive part of the field dynamics can be systematically derived using the Poisson-bracket (PB) formalisms, which can be generically written as~\cite{chaikin1995,stark2003,markovich2024}:
\begin{align}\label{eq:PB integral}
\hat{\mathcal{T}}\Psi_a(\bm{X}) 
&= 
-\int d\bm{X} \sum_b \lbrace \Psi_a(\bm{X}), \Psi_b(\bm{X}') \rbrace \frac{\delta H (\bm{X}')}{\delta \Psi_b(\bm{X}')} 
\, .
\end{align} 
Here $\bm{X}$ is the coordinate ($\bm{X}=\bm{r}$ for the Lagrangian description while $\bm{X}=\bm{R}$ for the Eulerian one). $\Psi_a$ refers to the CG field $a$, and $\delta H/\delta(\cdots)$ is the functional derivative of the Hamiltonian. The form of the time-evolution operator $\hat{\mathcal{T}}$ depends on the framework 
used. In the Lagrangian description, a pack of particles within the CG volume is tracked, leading to the \textit{total} time derivative for the evolution $\hat{\mathcal{T}}\Psi(\bm{r})=d\Psi(\bm{r})/dt$. In the Eulerian case, the local time derivative is used instead  $\hat{\mathcal{T}}\Psi(\bm{R})=\partial\Psi(\bm{R})/\partial t \equiv \dot \Psi (\bm{R})$, as we observe the particles at a fixed position in the real/deformed space.

The PB $\{\Psi_a(\bm{X}), \Psi_b(\bm{X}')\}$ in Eq.~\eqref{eq:PB integral} determines the dynamic coupling between two CG fields at positions $\bm{X}$ and $\bm{X}'$. It provides the bridge from \textit{microscopic particle} dynamics to \textit{macroscopic field} dynamics through
\begin{align}\label{eq:PB expression}
\lbrace \Psi_a(\bm{X}), \Psi_b(\bm{X}') \rbrace 
\equiv 
\sum_{\alpha, k}\bigg[ 
\frac{\partial\Psi_a(\bm{X})}{\partial \pi_k^\alpha}\frac{\partial\Psi_b(\bm{X}')}{\partial q_k^\alpha}-
\frac{\partial\Psi_a(\bm{X})}{\partial q_k^\alpha}\frac{\partial\Psi_b(\bm{X}')}{\partial \pi_k^\alpha}
\bigg]
\, .
\end{align}
Here $\pi_k^\alpha$ and $q_k^\alpha$ denote the canonical conjugates of the generalized microscopic momenta and positions of particle $\alpha$, respectively, and $k$ is the coordinate index. These pairs $(\pi_k^\alpha,q_k^\alpha)$ are defined in the \textit{real/Eulerian} space, which is the natural space for formulating the Hamiltonian dynamics and thus forms the foundation of the PB formalism. For instance, the canonical conjugate of the Eulerian particle position $\bm{R}^\alpha$ is its CM momentum $\bm{P}^\alpha$, whereas for the 2D particle internal rotation $\theta^\alpha$ used in the micropolar/Cosserat materials, it is the particle angular momentum $\ell^\alpha$. However, note that this rotational conjugacy does not hold for the 3D case, see Appendix~\ref{app:canonical conjugate pair}. 
The PBs are calculated using the fields as defined {in Eqs.~\eqref{eq:Lagran density fields}, \eqref{eq:Eulerian density fields} and \eqref{eq:particle-averaged fields}} in the previous section, {{which are not yet coarse-grained. The result of Eq.~\eqref{eq:PB expression} is then  coarse-grained and substituted in Eq.~\eqref{eq:PB integral} (for details on the CG procedure, see \cite{markovich2024}).}}

Importantly, {the fact that the canonical pairs are defined in the Eulerian description underlies the  differences between the Eulerian and Lagrangian PBs}. Specifically, when calculating the spatial derivative of a field, $\partial\bm{\Psi}/\partial \bm{q}^\alpha$, many Lagrangian PBs vanish because $\partial \delta(\bm{r}-\bm{r}^\alpha)/\partial \bm{R}^\alpha=0$, reflecting the essence of the Lagrangian formalism, where the same particle pack is tracked. 
%
%
In contrast, Eulerian PBs involving the CM momentum density, i.e., $\{ g^c(\bm{R}), \Psi_a(\bm{R}')\}$, generally contain additional, higher-order contributions because $\partial \delta(\bm{R}'-\bm{R}^\alpha)/\partial R^\alpha_i = -\nabla'_i \delta(\bm{R}'-\bm{R}^\alpha)$, with $\nabla'_i \equiv \partial / \partial R'_i$. This non-vanishing derivative physically accounts for the exchange of particles between neighboring CG volumes due to particle flow and material deformation (Fig.~\ref{fig:field wrt volume change}(b)). 

Consider, as an illustrative example, the PBs between the CM momenta, $\bm{g}^\circ$ (Lagrangian) and $\bm{g}^c$ (Eulerian), and the displacement $\bm{u}$ (detailed derivation is in Appendix~\ref{app:Eulerian PBs}): 
%
\begin{align}
\lbrace u_i(\bm{r}),  g_j^\circ(\bm{r}') \rbrace   
=& 
-\delta_{ij}\delta(\bm{r}-\bm{r}')
\, ,
\label{eq:gc u Lagran PBs}
\\  
\lbrace u_i(\bm{R}),g_j^c(\bm{R}') \rbrace   
=& 
-\big[\delta_{ij}- \nabla_j' u_i(\bm{R})\big]\delta(\bm{R}'-\bm{R})
\, .
\label{eq:gc u Eule PBs}
\end{align}
The common Kronecker-delta term $\delta_{ij}$ comes from differentiating the \textit{particle (microscopic)} displacement, $\bm{u}^\alpha=\bm{R}^\alpha-\bm{r}^\alpha$, with respect to the particle position $\bm{R}^\alpha$. In the Eulerian framework (Eq.~\eqref{eq:gc u Eule PBs}), the additional term $[\nabla_j' u_i(\bm{R}')]\delta(\bm{R}-\bm{R}')$ appears from differentiating the displacement field, $\partial \big[\sum_\alpha u^\alpha_i \delta(\bm{R}'-\bm{R}^\alpha)/n(\bm{R}')\big]/\partial R^\alpha_j$. 
Using the PB formalism (Eq.~\eqref{eq:PB integral}) for the displacement dynamics $\dot{u}_i$, this additional term then yields the convective contribution $v_j^c\nabla_j u_i$ when combined with $\delta H/\delta g^c_j(\bm{R}')=g^c_j(\bm{R}')/\rho(\bm{R}')\equiv v^c_j(\bm{R}')$, whereas the common $\delta_{ij}$ term gives the velocity contribution $v^c_i$. Accordingly, the Lagrangian and Eulerian descriptions of $\dot{\bm u}$ become, respectively,
\begin{align}\label{eq:u dynamics}
\frac{d u_i}{dt} = v^c_i \;\; ; \;\; \dot{u}_i + v^c_j\nabla_j u_i= v^c_i \,.
\end{align} 
From Eq.~\eqref{eq:u dynamics}, one can see that this difference between the Lagrangian and Eulerian PBs also leads to the transformation of the total time derivative: $d{\bm \Psi}/dt= \dot{\bm \Psi} + {\bm v}^c \cdot \nabla {\bm \Psi}$, as expected in continuum dynamics~\cite{LLelastity,eremeyev2013,LLfluidMechanics}. 
In Appendix~\ref{app:Eulerian PBs} we list the frequently-used Eulerian and Lagrangian PBs in isotropic elasticity with  particle internal rotations and also extend it to the 3D case.

The PBs give only  the reactive contribution to the dynamics~\cite{chaikin1995}. Unlike the microscopic dynamics, which is Hamiltonian, the CG procedure neglects many degrees of freedom, which appear as an additional dissipative term in the coarse-grained dynamics,
\begin{equation}
    \hat{\mathcal{T}} \Psi_a(\bm{X})\Big|_{\rm dissipative} = - \int d\bm{X}' \,  \Gamma_{ab}\left(\bm{X},\bm{X}'\right) \frac{\delta H}{\delta \Psi_b(\bm{X}')} \, ,
\end{equation}
such that the complete CG dynamics is:
\begin{equation}
\hat{\mathcal{T}}\Psi_a(\bm{X}) 
= 
-\int d\bm{X} \sum_b \lbrace \Psi_a(\bm{X}), \Psi_b(\bm{X}') \rbrace \frac{\delta H (\bm{X}')}{\delta \Psi_b(\bm{X}')} 
- \int d\bm{X}' \,  \Gamma_{ab}\left(\bm{X},\bm{X}'\right) \frac{\delta H}{\delta \Psi_b(\bm{X}')} \, .
\end{equation}
The dissipative contribution cannot be derived using PBs, and is usually derived using symmetry arguments (see, e.g., \cite{lubensky2005}). The dissipative tensor $\bm{\Gamma}$ is symmetric and positive semidefinite. It is generally a function of all fields, and following Curie’s symmetry principle, it must obey the system symmetries. Moreover, $\Psi_a$ is dissipatively coupled to  $\Psi_b$ only if they have the same sign under time reversal. This is the signature of dissipation where the flux $\partial \Psi_a/\partial t$ has the opposite sign under time reversal from the force $-\delta H/\delta \Psi_b$.
In this work we do not account for dissipation, thus, these terms will be ignored hereafter.

\subsection{Transformation of field dynamics}\label{sec:dynamic transformation}

Sections~\ref{sec:CG fields}--\ref{sec:PBs} develop the PB formalisms used to derive the Eulerian and Lagrangian field dynamics from the continuum Hamiltonian and the definitions of the CG fields. To show that these two PB frameworks are mutually consistent, and thereby complete the procedural loop shown in Fig.~\ref{fig:method}, we now provide a direct method for transforming the derived dynamic equations from one framework to the other. For this purpose, the CG fields must first be classified as either density-type fields or particle-averaged fields, according to their definitions in Sec.~\ref{sec:CG fields}.
 
For particle-averaged fields (Eq.~\eqref{eq:particle-averaged fields}), the transformation of the dynamics is straightforward. One simply writes the total time derivative in Eulerian form as $d{\bm \Psi}/dt=\dot{\bm \Psi}+{\bm v}^c\cdot\nabla{\bm \Psi}$. This relation, for instance, easily recovers the conversion between the two equations of the displacement dynamics previously derived by the two PB formalisms in Eq.~\eqref{eq:u dynamics}. 

For density-type fields, the transformation must also account for the volume change between the Lagrangian and Eulerian spaces. Applying the total time derivative $d/dt$ to both sides of the field transformation in Eq.~\eqref{eq:density field transformation} gives
\begin{align}
\frac{d \bm\Psi^\circ}{dt}
&=
\frac{d J}{dt} \bm\Psi
+
J
\big(
\dot{\bm\Psi} 
+ 
v_j^c\nabla_j\bm\Psi
\big)
=
J
\big[
\dot{\bm\Psi} +  \nabla_j(v^c_j \bm\Psi)
\big]
\, ,
\label{eq:dynamic transform E to L}
\end{align}
where we used the identity $(dJ/dt)/J= d\ln J/dt =\text{Tr}(\bm{F}^{-1}\dot{\bm{F}})= \nabla\cdot\bm v^c$~\footnote{{{Writing ${\bm F}(t+dt)\approx {\bm F}(t)+dt\,\dot{\bm F}$ gives $J(t+dt)\approx \det[\bm F(\bm I+dt\,\bm F^{-1}\dot{\bm F})]$. Taking the logarithm, using $\ln\det\bm m=\operatorname{Tr}\ln\bm{m}$, and expanding to first order in $dt$ yields $d\ln J/dt=\operatorname{Tr}(\bm F^{-1}\dot{\bm F})=\operatorname{Tr}[\bm F^{-1}(\partial\bm v^c/\partial\bm R)\bm F]=\nabla\cdot\bm v^c$.
}}}
%
%
to recover the streaming term $\nabla_j(v^c_j\bm\Psi)$. Thus, to transform Eulerian dynamics into its Lagrangian form, one first derives the combination $\dot{\bm\Psi}+\nabla_j(v^c_j\bm\Psi)$ from the Eulerian PB formalism and substitutes it into Eq.~\eqref{eq:dynamic transform E to L}, together with the derivative change $\nabla_i = F_{ji}^{-1} \nabla^\circ_j$. Conversely, starting from the explicit expression for $d\bm\Psi^\circ/dt$ obtained through the Lagrangian PB formalism, one can multiply both sides  by $J^{-1}$ and use $\nabla_i^\circ = F_{ji} \nabla_j$ to derive the corresponding Eulerian dynamics. Note that the factor $J$ generates higher-order terms in $\nabla^\circ_j u_i$ or $\nabla_j u_i$. In linear elasticity, this transformation of dynamics 
reduces to $d \bm{\Psi}^\circ/dt = \dot{\bm\Psi}+\nabla_j(v_j^c\bm\Psi)$. The difference between the Lagrangian and Eulerian dynamics then simply lies in replacing the total time derivative by the partial time derivative plus a streaming term.

The transformation of the dynamic equations in Eq.~\eqref{eq:dynamic transform E to L} recovers the standard relation between the Cauchy stress $\bm\sigma$ and the first Piola--Kirchhoff (1st PK) stress $\bm\sigma^\text{PK}$~\cite{LLelastity}. First, note that the Lagrangian and Eulerian PB formalisms naturally lead to these two different stress measures in the CM momentum dynamics:
\begin{align}
\frac{d g^\circ_i}{dt} 
= \nabla^\circ_j \sigma^\text{PK}_{ij}
\;\; ; \;\; 
\dot{g}^c_i  + \nabla_j(v_j^c g^c_i)
= \nabla_j \sigma_{ij}
\, .
\label{eq:stress equations}
\end{align}
Combining Eq.~\eqref{eq:dynamic transform E to L} with the Eulerian dynamics in Eq.~\eqref{eq:stress equations} then reveals 
\begin{align}
\frac{d g_i^\circ}{dt}
&=
J \nabla_l \sigma_{il}
= J F_{jl}^{-1}\nabla_j^\circ\sigma_{il}
= \nabla^\circ_j
\big(
\underbrace{
J \sigma_{il} F_{jl}^{-1}}_{\equiv \sigma^\text{PK}_{ij}} 
\big)
\, ,
\end{align}
where we used the Piola identity $\nabla_j^\circ(JF^{-1}_{jl})=0$~\cite{gurtin1981}.


%


%
\begin{figure}[t]
	\centering
	\includegraphics[width=14
cm]{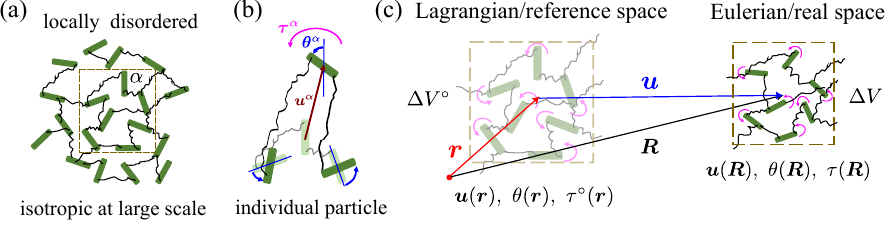}
\caption{(a) Model of the isotropic, disordered chiral active solid studied here, represented as a network of interconnected rod-like particles in the spirit of Cosserat elasticity theory. (b) Each particle $\alpha$ has a center-of-mass (CM) displacement $\bm{u}^\alpha$ and can internally rotate by an angle $\theta^\alpha$ away from its equilibrium orientation, indicated by the blue line. This internal rotation can be driven by an active torque $\tau^\alpha$ acting on the particle as the chiral input. (c) Coarse-graining (CG) schemes in the Lagrangian (left) and Eulerian (right) frameworks, using the coordinates of the undeformed/reference space $\bm{r}$ and the deformed/real space $\bm{R}$, respectively. The two coordinates are related by the displacement field $\bm{u}$ through $\bm{R}=\bm{r}+\bm{u}$. The displacement field $\bm{u}$ and the internal rotation field $\theta$ are particle-averaged fields, as defined in Eq.~\eqref{eq:particle-averaged fields}, and therefore their values are unchanged when switching between the two frameworks. By contrast, the active torque density is obtained by summing the particle contributions within a CG volume, and is therefore affected by the CG volume change: $\tau^\circ(\bm{r})\Delta V^\circ=\tau(\bm{R})\Delta V$.}
	\label{fig:model}
\end{figure}

\section{Chiral odd solids -- a minimal example}\label{sec:example}

In Secs.~\ref{sec:field H} and \ref{sec:PBs}, we showed that the Eulerian framework generates higher-order contributions both through the transformation of the Hamiltonian into Eulerian form and through the Eulerian PBs. These terms  become important in the presence of large deformations or non-linear stress-strain relations.
%
Here we focus on the former, where large (non-linear) deformations are driven by external fields or by internal active forces or torques. {{Importantly, the stress-strain relation remains linear in what follows.}}
In this section we  illustrate this by using a minimal model of chiral active solids, where active torques imposed at the particle level provide the chiral input and select a preferred rotational direction, while also causing non-linear deformations. 

To this end, we consider a 2D solid composed of identical, finite-sized particles. This solid is assumed to be homogeneous and isotropic at large scales, but locally disordered, as in biological gels such as the cytoskeleton~\cite{Broedersz2014} and in synthesized materials such as elastomers with embedded magnetic colloids~\cite{lucarini2022,moreno-mateos2022}. 
The particles are further treated as rigid bodies, allowing only translation of the CM and rotation around the CM, namely, $\bm{u}^\alpha$ and $\theta^\alpha$ in Fig.~\ref{fig:model}, with $\alpha$ being the particle index. We incorporate the elastic effects of these internal rotations $\theta^\alpha$ in the spirit of micropolar (Cosserat) elasticity~\cite{eringen1966,eringen1999,eremeyev2013}. Compared with classical elasticity~\cite{LLelastity}, this micropolar description introduces an additional rotational degree of freedom per particle, therefore giving rise to an additional deformation field and an additional strain~\cite{eringen1999,eremeyev2013,lee2026a} (see Sec.~\ref{sec:potential}).

Local active torques $\tau^\alpha$, imposed, for example, by actomyosin motor proteins or an external magnetic field, drive internal particle rotations, which may result in \textit{large} angular change. These large rotations induce \textit{geometric nonlinearities}, which must be accounted for in the elastic potential. 


\subsection{Lagrangian elastic potential}\label{sec:potential}\label{sec:potential}

Since torques do work though internal particle rotations, the Lagrangian total potential density $V^\circ_\text{tot}$ includes, in addition to an elastic potential density $V^\circ$, an active-torque contribution, $-\tau^\circ\theta$~\footnote{The torque applied per particle gives a contribution to the potential $-\sum_{\alpha \in \Delta V^\circ} \tau^\alpha \theta^\alpha \delta(\bm{r}-\bm{r}^\alpha) \approx -\sum_{\alpha \in \Delta V^\circ} \tau^\alpha \delta(\bm{r}-\bm{r}^\alpha)\sum_{\beta \in \Delta V^\circ}  \theta^\beta \delta(\bm{r}-\bm{r}^\beta) / n^\circ(\bm{r})$~\cite{markovich2024}, which after coarse-graining becomes $-\tau^\circ(\bm{r})\theta(\bm{r})$.}:
\begin{align}\label{eq:V total}
V^\circ_\text{tot}=
- \tau^\circ\theta   
+ V^\circ 
\, . 
\end{align}
This active torque potential density is similar to the potential due to an external field, although no global alignment is present here. It naturally introduces the active torque density $\tau^\circ$ into the angular momentum dynamics $\ell^\circ$ through the Lagrangian PB formalism~\cite{lee2026a}.


To formulate the elastic potential density $V^\circ$, we first describe strains using the particle-averaged fields of displacement $\bm{u}$ and internal rotation $\theta$ (see definitions in Eq.~\eqref{eq:particle-averaged fields}). Since both $\bm{u}$ and $\theta$ cause deformations, this leads to the Green-Lagrange strain $u_{ij} \equiv
\big(F_{ki}F_{kj}-\delta_{ij} \big)/2$ and another strain that couples the displacement with the internal rotation $e_{ij}
\equiv O_{ki}F_{kj}-\delta_{ij}$~\cite{eringen1999,eremeyev2013,lee2026a}. Here $O_{ik}=\delta_{ik}-\varepsilon_{ik} \sin\theta- (1-\cos\theta)\delta_{ik}$ is the 2D rotation matrix, with $\varepsilon_{ij}$ being the 2D Levi-Civita symbol. In principle, the strains can also depend on $\nabla^\circ_j\theta$~\cite{chaikin1995,eringen1999,eremeyev2013,lee2026a}. However, in the hydrodynamic limit, such terms  are negligible and ignored hereafter. This is especially appropriate for systems of small particles interconnected over a relatively large separation length, such that the rotation of an individual particle has only a weak effect on its neighbors~\cite{lee2026a}.






Assuming linear stress-strain relations for these two strains, we formulate the elastic potential density $V^\circ$ as a quadratic form (see Ref.~\cite{lee2026a} for details). We further assume small displacement, while retaining the \textit{geometric nonlinearities} arising from active-torque driven rotations $\theta$. Specifically, we focus on the regime $|\nabla_i^\circ u_j|\ll \theta \ll 1$, where the internal rotation remains small but larger than  the displacement gradient. This way, we expand $V^\circ$ only to the lowest-order non-linear contribution~\cite{lee2026a}:
\begin{align}
V^\circ
=&
\frac{E_{ijkl}}{2}
u^s_{ij} u^s_{kl} 
+ 
\kappa_c (\phi^{\circ})^2 
-(\tilde\lambda+\tilde\mu-\kappa_c)(\phi^{\circ})^2 (\nabla^\circ\cdot\bm{u})
\notag\\
&
+\big[
\tilde\mu
\big(
\varepsilon_{ij}\delta_{kl}
+
\varepsilon_{jk}\delta_{il}
\big)
-
\frac{\kappa_c}{2}\varepsilon_{ij}\delta_{kl}
\big]
\phi^\circ (\nabla^\circ_j u_i)(\nabla^\circ_l u_k)
\, .
\label{eq:Lagran V expression}
\end{align}
Here $u^s_{ij}\equiv(\nabla^\circ_j u_i+\nabla^\circ_i u_j)/2$ is the linearized Green-Lagrange strain with the symmetric elastic tensor $E_{ijkl}\equiv\lambda\delta_{ij}\delta_{kl} + \mu (\delta_{ik}\delta_{jl}+\delta_{il}\delta_{jk})$,
and the Lam\'{e} coefficients $\lambda$ and $\mu$ of the classical linear isotropic elasticity~\cite{LLelastity}. $\phi^\circ \equiv \theta-(\nabla^\circ\times\bm{u}/2)$ is the mismatch between the internal rotation and CM orbital rotation~\cite{eringen1999,eremeyev2013,lee2026a}. Note that $\phi^\circ$ is not a density field, and its superscript $\circ$ is used to distinguish from its Eulerian counterpart $\phi$ in Eq.~\eqref{eq:VE}, since their spatial gradients are defined in different coordinates.   

The first two terms in Eq.~\eqref{eq:Lagran V expression} correspond to the linear isotropic Cosserat elasticity with $\kappa_c$ denoting the Cosserat coupling constant~\cite{eringen1964,neff2006,eremeyev2013}. The remaining terms are the leading geometric nonlinearities involving $\phi^\circ$, where  $\tilde\lambda$ and $\tilde\mu$ are constant elastic coefficients. Note that since $\theta$ (equivalently, $\phi^\circ$) is an internal degree of freedom, it typically relaxes fast~\cite{chaikin1995,markovich2024,markovich2025,maitra2019,surowka2023} to the value determined by the local torque balance, $\partial V^\circ_\text{tot}/\partial\theta=0$. As a consequence, the rotational mismatch is linearly proportional to the active torque density $\phi^\circ\propto \tau^\circ$~\cite{lee2026a}. It is therefore expected that these nonlinear terms, which are proportional to the rotational mismatch $\phi^\circ$, become important only when active torque density $\tau^\circ$ is present~\footnote{In the absence of active torques $|\phi^\circ|\sim \mathcal{O}(|\nabla_j^\circ u_i|)$. The nonlinear terms are thus of the order $\mathcal{O}(|\nabla_j^\circ u_i|^3)$, which must be  discarded under the small-displacement assumption.}.

 


\subsection{Eulerian elastic potential}\label{sec:potential transformation}

In our previous work~\cite{lee2026a}, we applied the Lagrangian PB formalism to the Hamiltonian combining Eqs.~\eqref{eq:H} and \eqref{eq:Lagran V expression}. Then, we applied the stress transformation to find the Cauchy stress, which is the stress in the real (deformed) space where balance laws must hold (e.g., balance of angular momentum) and that is usually what is measured in experiments~\cite{veenstra2025a,tan2022}. Remarkably, odd elasticity~\cite{scheibner2020,braverman2021,fossati2024,fruchart2023,banerjee2025,lee2026a} naturally emerged from the geometric nonlinearities induced by active torques. 

To complete the procedural loop and demonstrate the consistency between the Lagrangian and Eulerian frameworks (Fig.~\ref{fig:method}), we will apply the Eulerian PB formalism directly to the Eulerian Hamiltonian (Eq.~\eqref{eq:H Eulerian}) and show that it recovers the same (Cauchy) stress tensor and the odd elastic modulus as in Ref.~\cite{lee2026a}. 


We start by transforming the Lagrangian total potential density $V^\circ_\text{tot}$ (Eq.~\eqref{eq:V total}) into its Eulerian form using Eq.~\eqref{eq:H Eulerian}. Since $\tau^\circ$ is a density-type field, the CG volume change is absorbed into the Eulerian torque density, $\tau=J^{-1}\tau^\circ$. The active potential density then takes the simple form $-\tau\theta$, and the Eulerian total potential density $V_\text{tot}$ can be written as:
\begin{equation}\label{eq:Eulerian Vtot}
V_\text{tot} 
=
-\tau\theta + V
\, ,
\end{equation}
where $V$ is the Eulerian elastic potential density. For the elastic part $V$, retaining geometric nonlinearities generates additional contributions in the Eulerian form. Specifically, transforming $(\phi^\circ)^2$ gives $J^{-1}(\phi^\circ)^2\approx(1-\nabla\cdot\bm{u})\phi^2-\varepsilon_{li}(\nabla_l u_k)(\nabla_k u_i)\phi$, where $\phi\equiv \theta-(\nabla\times\bm{u})/2$ is the Eulerian rotational mismatch. Here we have used $\nabla^\circ_i\equiv F_{ji}\nabla_j\approx(\delta_{ij}+\nabla_i u_j)\nabla_j$. The \textit{Eulerian} elastic potential density $V$ then reads:
\begin{align}\label{eq:VE}
V
=&
\frac{E_{ijkl}}{2}
\bar{u}^s_{ij} \bar{u}^s_{kl} 
+ 
\kappa_c \phi^2 
-
(\tilde\lambda+\tilde\mu)\phi^2(\nabla\cdot\bm{u})
\notag\\
&
+\big[
\tilde\mu
\big(
\varepsilon_{ij}\delta_{kl}
+
\varepsilon_{jk}\delta_{il}
\big)
-\kappa_c
\big(
\frac{1}{2}\varepsilon_{ij}\delta_{kl}
+\varepsilon_{li}\delta_{jk}
\big)
\big]
\phi (\nabla_j u_i)(\nabla_l u_k)
\, ,
\end{align}
where $\bar{u}^s_{ij}\equiv (\nabla_j u_i +\nabla_i u_j)/2$ is the linear symmetrized Eulerian strain. 

Although the Eulerian rotational mismatch $\phi$ differs from its Lagrangian counterpart $\phi^\circ$ in that it uses $\nabla$ instead of $\nabla^\circ$, it is driven by active torques in the same way. 
Fast relaxation of the internal rotation, $\partial V/\partial\theta=0$, gives $\phi\propto\tau$, analogous to the Lagrangian relation $\phi^\circ\propto\tau^\circ$. Thus, without active torques, the nonlinear terms vanish and the Eulerian elastic potential reduces to the linear isotropic Cosserat form (i.e., the first two terms in $V$), as in the Lagrangian framework. 
The difference between the Lagrangian and Eulerian elastic potentials  appears only in the coefficients of the nonlinear terms $\phi^2(\nabla\cdot\bm{u})$ and $\phi(\nabla_j u_i)(\nabla_l u_k)$ (compare Eqs.~\eqref{eq:Lagran V expression} and \eqref{eq:VE}). 

\subsection{Eulerian PB formalism -- dynamics, stress tensor, and odd elasticity}\label{sec:stress and odd elasticity}

We now apply the Eulerian PB formalism (Eqs.~\eqref{eq:H Eulerian}--\eqref{eq:PB expression}) to the Hamiltonian (Eqs.~\eqref{eq:Eulerian Vtot} and \eqref{eq:VE}) and derive the dynamics of the CM momentum and angular momentum. For the potential contributions, we require the following non-vanishing Eulerian PBs involving $\bm{u}$, $\theta$, and $\tau$:
\begin{align}
\lbrace g_i^c(\bm{R}), u_j(\bm{R}') \rbrace   
=& 
\big[\delta_{ij}- \nabla_j' u_i(\bm{R})\big]\delta(\bm{R}-\bm{R}')
\label{eq:gc u PB}
\, ,
\\
\lbrace g_i^c(\bm{R}), \theta(\bm{R}') \rbrace    
=& 
-\big[\nabla_i' \theta (\bm{R}') \big]\delta(\bm{R}-\bm{R}')
\label{eq:gc theta PB}
\, ,
\\
\lbrace g_i^c(\bm{R}),\tau(\bm{R}') \rbrace
=&
- \nabla_i'\big[
\tau(\bm{R}')\delta(\bm{R}-\bm{R}')
\big]
\label{eq:gc tau PB}
\, ,
\\
\lbrace \ell(\bm{R}),\theta(\bm{R}') \rbrace
=&
\delta(\bm{R}-\bm{R}')
\label{eq:ell theta PB}
\, . 
\end{align}
These PBs provide the mechanical part of the dynamics, such as stress and torque terms. The remaining PBs, not listed here, involve fields that  enter only the kinetic energy and give the usual streaming terms~\cite{stark2003,stark2005,markovich2021,markovich2024,lee2026a}. We also note that, in the present 2D example, $\ell$ and $\theta$ form a canonical pair inherited from the microscopic conjugacy between $\ell^\alpha$ and $\theta^\alpha$ (see Appendix~\ref{app:canonical conjugate pair}), yet this simple canonical structure does not extend directly to 3D. In Appendix~\ref{app:Eulerian PBs} we derive the complete list of PBs, together with their 3D generalization.

Substituting the PBs of Eqs.~\eqref{eq:gc u PB}--\eqref{eq:ell theta PB} into Eq.~\eqref{eq:PB integral}, we obtain the balance equations for the CM- and angular-momentum:
\begin{align}
\notag\dot{g}^c_i
+\nabla_j(v^c_j g^c_i)
&=
\frac{\delta H}{\delta u_j}
(
\nabla_i u_j
-\delta_{ij}
)
+ 
\big(
\frac{\delta H}{\delta \theta}+\tau
\big)
\nabla_i\theta
\\
&=
\nabla_j
\big\lbrace
E_{ijkl} \nabla_l u_k
+ \kappa_c\varepsilon_{ij}\phi
-\phi^2(\tilde\lambda+\tilde\mu-\kappa_c)\delta_{ij}
\notag\\
&\hspace{75pt}
+\phi\big[
\frac{\kappa_c}{2}(\varepsilon_{ik}\delta_{jl}+\varepsilon_{jl}\delta_{ik})
-(\tilde\lambda+\tilde\mu)\varepsilon_{ij}\delta_{kl}
\big]\nabla_l u_k
\big\rbrace
\, ,
\label{eq:gc dynamics}
\\
\notag\dot{\ell}
+\nabla_j(v_j^c\ell)
&=
-\frac{\delta H}{\delta \theta} 
= \tau -\frac{\partial V}{\partial \phi}
\\
&=
\tau
-
2 \kappa_c 
\phi
+
2(\lambda+\mu)\phi (\nabla\cdot\bm{u})
\label{eq:Eulerian ell dynamics}
\, .
\end{align}
Compared with the Lagrangian equations $d g^\circ_i/dt=-\delta H/\delta u_i$ and $d\ell^\circ/dt=-\delta H/\delta\theta$~\cite{stenull2004,lee2026a}, the Eulerian CM momentum equation $\dot{\bm{g}}^c$ contains additional spatial-derivative terms generated by the Eulerian PBs. These terms reflect the exchange of particles between neighboring CG volumes (see also the discussion in Sec.~\ref{sec:PBs}). They produce the nonlinear contributions to the Cauchy stress, shown inside the braces in Eq.~\eqref{eq:gc dynamics}, and are crucial for the emergence of the odd elasticity
(see the complete derivation in Appendix~\ref{app:2D dynamic eqs}). By contrast, the angular momentum equation retains essentially the same structure $-\delta H /\delta \theta$ in the two frameworks.

Since $\theta$ (or equivalently $\phi$) is an internal degree of freedom~\cite{chaikin1995,markovich2024,markovich2025,maitra2019,surowka2023}, it typically relaxes fast compared with the hydrodynamic displacement field $\bm{u}$, such that $\partial H/\partial \theta = 0$. This corresponds to the mechanical balance $\dot{\ell}=0$ and $\ell=0$ in Eq.~\eqref{eq:Eulerian ell dynamics}, which results in internal rotation:
\begin{align}\label{eq:fast angle relax}
\phi 
\approx
\frac{\tau}{2\kappa_c}
\bigg(
1+\frac{\tilde\lambda+\tilde\mu}{\kappa_c}\nabla\cdot\bm{u}
\bigg)
\, ,
\end{align}
where we have used $(\tilde\lambda+\tilde\mu)(\nabla\cdot\bm{u})/\kappa_c \ll 1$ to neglect higher-order terms.

Substituting Eq.~\eqref{eq:fast angle relax} into Eq.~\eqref{eq:gc dynamics}
we get
\begin{align}
&\dot{g}^c_i + \nabla_j(v^c_j g^c_i)
=
\nabla_j
\big\lbrace
\underbrace{
\frac{\tau}{2}\varepsilon_{ij}
-\frac{\tau^2}{4\kappa_c}(\tilde\lambda+\tilde\mu-\kappa_c)\delta_{ij}
}_{\bm\sigma^\text{pre}}
+
\big[
\underbrace{
 E_{ijkl}
 +
\frac{\tau}{4}(\varepsilon_{ik}\delta_{jl}+\varepsilon_{jl}\delta_{ik})
}_{\bm{C}}
\big]
\nabla_l u_k
\big\rbrace
\, .
\label{eq:gc Eul relaxsed dynamics}
\end{align}
Here $\bm\sigma^\text{pre}$ is the active-torque induced prestress that appears even in the absence of deformation. $\bm{C}$ is the real-space elasticity tensor, which, when written in the orthogonal basis of  deformations/stresses~\footnote{Expressing the elasticity tensor 
$C_{ijkl} = \big[
B\delta_{ij}\delta_{kl}
+
\mu(\delta_{ik}\delta_{jl}+\delta_{il}\delta_{jk}-\delta_{ij}\delta_{kl})
+K^o(\varepsilon_{ik}\delta_{jl}+\varepsilon_{jl}\delta_{ik})
\big]$ in the 2D irreducible basis of deformations/stresses, gives  Eq.~\eqref{eq:elasticity tensor}~\cite{scheibner2020,fruchart2023}.}, takes the following form:
\begin{equation}\label{eq:elasticity tensor}
\begin{matrix}
\includegraphics[width=.45\textwidth]{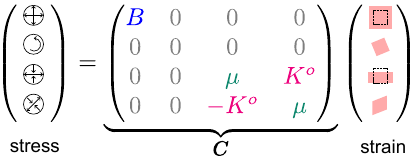}
\end{matrix}
\ \ .
\end{equation}
The bulk modulus $B=\lambda+\mu$, and the odd elastic modulus is proportional to the active torques $K^o = \tau/4$.

As required, Eqs.~\eqref{eq:gc Eul relaxsed dynamics} and \eqref{eq:elasticity tensor} recover the stress and elasticity tensors of Ref.~\cite{lee2026a} in which the Lagrangian PB formalism was applied together with the stress transformation.


\section{Conclusion}\label{sec:conclusion}

In this work, we have developed a systematic procedure for applying the Eulerian PB formalism to elastic potentials, which are commonly formulated in Lagrangian space. By combining the transformation of the Hamiltonian, the transformation of derived dynamic equations, and the relation between stress measures, we completed the procedural loop connecting the Lagrangian and Eulerian PB formalisms (Fig.~\ref{fig:method}) and generically demonstrated their consistency.

We further clarified the key differences between the two formalisms. In the Eulerian framework, additional nonlinear terms arise from two sources. The first comes from transforming the Hamiltonian from the Lagrangian coordinate $\bm{r}$ to the Eulerian coordinate $\bm{R}$, where the CG volume change introduces factors of $J^{-1}$ and the spatial derivatives must be rewritten through $\nabla^\circ_i=F_{ji}\nabla_j$. The second comes from Eulerian PBs involving the CM momentum density $\bm{g}^c$, which account for particle exchange between neighboring CG volumes. Mathematically, this second contribution originates from the non-vanishing derivative  acting on the CG fields $\partial\delta(\bm{R}-\bm{R}^\alpha)/\partial\bm{R}^\alpha$. This in turn generates additional spatial-derivative terms in the dynamic equations. 
In linear elasticity, these higher-order contributions are neglected, and the Lagrangian and Eulerian PB formalisms coincide.

However, when nonlinearities become important, these additional, higher-order Eulerian contributions must be taken into account to correctly capture the elastic response. We illustrated this point using chiral active solids as a minimal example~\cite{lee2026a}. In these systems, active torques 
drive particle internal rotations to induce the geometric nonlinearities that couple the rotational mismatch and displacement gradients~\cite{eringen1999,eremeyev2013,lee2026a}. In this example, nonlinearities must be included and are essential for the emergence of the odd elastic modulus~\cite{scheibner2020,fruchart2023}. This new elastic modulus  has attracted growing interest because of its non-reciprocal coupling between the two shear modes and the resultant unusual mechanical responses (see, e.g., Refs.~\cite{lee2026a} and \cite{veenstra2025}).

By recovering the odd elastic modulus directly from the Eulerian PB formalism, this work demonstrates that the Eulerian PB framework can consistently capture nonlinear elastic effects in the deformed (real) space. More broadly, our scheme  provides a route to study systems naturally formulated in real (Eulerian) space, and specifically driven active solids whose driving sources can be incorporated into a potential and whose experimentally measured stresses are naturally defined in real space~\cite{tan2022}. This opens a way to connect microscopic particle dynamics, such as force- or torque-driven internal motion, to emergent macroscopic elastic behavior.



\ack{This research was supported in part by Grant No. 2022/369 from the United States-Israel Binational Science Foundation (BSF). T.M. acknowledges funding from the Israel Science Foundation (Grant No. 1356/22).}





\appendix

\makeatletter
\renewcommand{\@seccntformat}[1]{%
  \ifstrequal{#1}{section}
    {Appendix~\thesection.\quad}
    {\csname the#1\endcsname\quad}}
\makeatother

\section{Canonical conjugate variables in 3D} \label{app:canonical conjugate pair}

The main text focuses on the 2D case. Here we derive the PB expressions for 3D systems. To this end, we first identify the canonical momenta conjugate to the \textit{real-space} particle $\alpha$ position $\bm{R}^\alpha$ and its vector of internal rotations $\bm\theta^\alpha$. We describe rotations using the axis-angle representation~\cite{bauchau2003,gallego2015}, which is common in micropolar (Cosserat) elasticity~\cite{eringen1999,eremeyev2020} (see more details below, above Eq.~\eqref{appeq:rotation}). 
%
These canonical momenta are defined from the Lagrangian $\mathcal{L}$, as $\partial\mathcal{L}/\partial\dot{\bm{R}}^\alpha$ and $\partial\mathcal{L}/\partial\dot{\bm{\theta}}^\alpha$~\cite{Goldstein_book}, where $\dot{\bm X}^\alpha\equiv d\bm X^\alpha/dt$ is the time derivative for particle quantities. 
The Lagrangian of the 3D system is:
\begin{align}\label{appeq:Lagrangian}
\mathcal{L}=
\sum_\alpha 
\left[
\frac{{(\bm{p}^\alpha})^2}{2 m^\alpha} 
+\frac{1}{2} \Omega_i^\alpha I_{ij}^\alpha \Omega_j^\alpha
\right]
- \mathcal{V}(\{ \bm{\theta^\alpha}, \bm{u}^\alpha \})
,   
\end{align}
where $\bm{p}^\alpha = m^\alpha \dot{\bm{R}}^\alpha$ is particle $\alpha$'s linear momentum and $\bm\Omega^\alpha$ is its angular velocity,
with $\ell_i^\alpha=I_{ij}^\alpha \Omega^\alpha_j$ being the particle angular momentum.
Here $m^\alpha$ and $\bm{I}^\alpha$ are the particle mass and moment of inertia, respectively. The sum in Eq.~\eqref{appeq:Lagrangian} is the kinetic energy, while $\mathcal{V}$ is the potential energy, which is only a function of  the particles' internal rotation $\bm\theta^\alpha$ and displacement $\bm{u}^\alpha$. Hence, $\partial \mathcal{L}/\partial \dot{\bm{R}}^\alpha$ and $\partial \mathcal{L}/\partial \dot{\bm\theta}^\alpha$ apply only to the kinetic part of $\mathcal{L}$. This gives the canonical conjugate pair $(\bm{R}^\alpha,\bm{p}^\alpha)$ for the particle position. For the internal rotation $\bm{\theta}^\alpha$, its  conjugate momentum reads
\begin{align}\label{appeq:ell tilde vs ell}
\tilde{\ell}^\alpha_i 
= \frac{\partial \mathcal{L} }{\partial \dot{\theta}_i^\alpha}
=
\frac{1}{2}\frac{\partial (I^\alpha_{jk}\Omega^\alpha_j\Omega^\alpha_k)}{\partial \dot{\theta}^\alpha_i}
= 
\frac{\partial \Omega_j^\alpha}{\partial\dot{\theta}^\alpha_i}
\ell_j^\alpha
\, .   
\end{align}
%

To evaluate $\partial\Omega_j^\alpha/\partial\dot{\theta}^\alpha_i$, 
we use the axis-angle representation for $\bm{\theta}^\alpha$. In this representation, a rotation is parametrized by a vector $\bm{\psi}$ (in our case, $\bm\psi = \bm\theta^\alpha$). Its direction $\bm{s}\equiv\bm{\psi}/\psi$ specifies the rotation axis, while its magnitude $|\bm{\psi}|\equiv\psi=(\psi_i\psi_i)^{1/2}$ gives the rotation angle. The corresponding rotation matrix is given by Rodrigues' formula~\cite{bauchau2003}:
\begin{align}\label{appeq:rotation}
\notag O_{ij} 
&
= 
\delta_{ij} 
+ \sin\psi\,(\bm{s}\times)_{ij} 
+(1-\cos\psi) (\bm{s}\times)_{ik}(\bm{s}\times)_{kj}
\\
&=
\delta_{ij} 
- \sin\psi\, \varepsilon_{ijk} \frac{\psi_k}{\psi}
+ (1-\cos\psi)\varepsilon_{ink}\varepsilon_{ksj} \frac{\psi_s\psi_n}{\psi^2}
\, ,
\end{align}
where $(\bm{s}\times)_{ij}\equiv-\varepsilon_{ijk}s_k=-\varepsilon_{ijk}\psi_k/\psi$ is the matrix representation of the cross product with $\bm{s}$. 

{To derive the angular velocity $\bm{\Omega}$, we consider a rigid reference vector $\bm{q}_0$ at some initial time, whose length is preserved under rotation. Its time-dependent orientation is given by $\bm{q}=\bm{O}\bm{q}_0$, where $\bm{O}$ is the rotation matrix (Eq.~\eqref{appeq:rotation}). Taking the time derivative gives $d\bm{q}/dt=(d\bm{O}/dt)\bm{q}_0$. Using $\bm{q}_0=\bm{O}^T\bm{q}$, we obtain $d\bm{q}/dt=(d\bm{O}/dt)\bm{O}^T\bm{q}= \bm\Omega\times\bm{q}$, where the last equality comes from the kinematics of a rigid body. The angular velocity can then be calculated from 
$\Omega_j=-\varepsilon_{jlk}(dO_{lm}/dt)O_{km}/2$, which yields~\cite{bauchau2003,gallego2015}:}
\begin{align}
\notag \Omega_j 
&= \dot\psi s_j%
+\big[ 
\sin\psi \dot{s}_j + (1-\cos\psi)\varepsilon_{jlk}s_l\dot{s}_k
\big]
\\
&=
\dot{\psi}\frac{\psi_j}{\psi}
+
\frac{\sin\psi}{\psi^2}
(\dot{\psi}_j \psi - \psi_j\dot{\psi})
+
\frac{1-\cos\psi}{\psi^3}\varepsilon_{jlk}\psi_l(\dot\psi_k\psi-\psi_k\dot\psi)
\label{appeq:angular velocity form 2}
\, .
\end{align}
The square-bracketed term in the first line of Eq.~\eqref{appeq:angular velocity form 2} arises from changes in the direction of the rotation axis. If the rotation axis is fixed, $\dot{\bm{s}}=0$, the angular velocity reduces to $\bm{\Omega}=\dot{\psi}\bm{s}$, as one would expect for 2D systems.



Direct differentiation of Eq.~\eqref{appeq:angular velocity form 2} with respect to $\dot{\psi}_i$ gives the matrix $\mathcal M^{-1}_{ij}$ that relates the angular momentum $\bm{\ell}$ to the canonical angular momentum $\tilde{\bm{\ell}}$ through $\tilde{\ell}_i=\mathcal M^{-1}_{ij}\ell_j $ (see Eq.~\eqref{appeq:ell tilde vs ell} with $\bm \psi = \bm\theta^\alpha$ and we suppress the particle index $\alpha$ for brevity): 
\begin{align}
\notag\frac{\partial \Omega_j}{\partial\dot{\psi}_i}
\equiv
\mathcal M_{ij}^{{-1}}
&
=
\delta_{ij}
+\bigg[
\frac{1-\cos\psi}{\psi^2}\psi_n \varepsilon_{jnk}
-(1-\frac{\sin\psi}{\psi})\delta_{jk}
\bigg]
\bigg(
\delta_{ki}- \frac{\psi_k\psi_i}{\psi^2}
\bigg)
\label{appeq:M-1}
\\
&
=
P^{\parallel}_{ij}
+
\frac{\sin\psi}{\psi}P^{\perp}_{ij}
-
\frac{1-\cos\psi}{\psi^2}A_{ij}
\, ,
\end{align}
where we define the symmetric parallel projector tensor $P^\parallel_{ij}\equiv \psi_i\psi_j/\psi^2$, the symmetric perpendicular projector $P^\perp_{ij}\equiv \delta_{ij} -( \psi_i\psi_j/\psi^2)$ and the antisymmetric tensor $A_{ij}\equiv -\varepsilon_{ijk} \psi_k$. Geometrically, using 
$\bm P^\parallel$ (or $\bm P^\perp$) on an arbitrary vector yields the parallel (or perpendicular) component along the rotational direction $\bm \psi$. These tensors have the following useful identities, which we use later to invert the matrix  $\bm{\mathcal{M}}^{-1}$: 
$\bm{P}^\parallel\bm{P}^\parallel=\bm{P}^\parallel$, $\bm{P}^\perp\bm{P}^\perp=\bm{P}^\perp$, $\bm P^\parallel \bm P^\perp = 0$, $\bm P^\parallel \bm A = \bm A \bm P^\parallel = 0$, $\bm P^\perp \bm A = \bm A \bm P^\perp = \bm A$, and $\bm A \bm A = -\psi^2  \bm P^\perp$. 

The matrix $\bm{\mathcal{M}}^{-1}$ in Eq.~\eqref{appeq:M-1} shows that the \textit{canonical} angular momenta $\tilde{\ell}_j$ for the rotation $\psi_j$ differ from the usual angular momenta $\ell_j$, reflecting the effects of changing the rotational direction. Since $\bm{\mathcal M}^{-1}$ depends only on the rotation vector $\bm{\psi}$, and not on $\tilde{\bm{\ell}}$, its inverse matrix $\bm{\mathcal M}$ gives the differential relation between $\bm\ell$ and $\tilde{\bm\ell}$ (see Eq.~\eqref{appeq:ell tilde vs ell}), which is useful for calculating PBs. Taking $\bm{\mathcal{M}} = \bm P^\parallel + C_\perp \bm P^\perp + C_A \bm A$ and using  $\mathcal M_{ik} \mathcal M^{-1}_{kj}=\delta_{ij} = P^\parallel_{ij} + P^\perp_{ij}$, one can determine the coefficients $C_\perp$ and $C_A$ and obtain
\begin{align}
\mathcal M_{ij}
&
=
\bigg(
\frac{\partial \ell_i}{\partial \tilde{\ell}_j}
\bigg)_{\bm{\psi}}
=
P^\parallel_{ij}
+
\bigg(
\frac{\psi}{2}\cot\frac{\psi}{2}
\bigg)
P^\perp_{ij}
+
\frac{1}{2}A_{ij}
\label{appeq:M}
\, .    
\end{align}

For 2D systems, $\mathcal{M}_{ij}^{-1}=\mathcal{M}_{ij}=\delta_{ij}$.
Thus, the canonical momentum conjugate to the 2D internal rotation is simply the usual angular momentum, as used in the main text. 

\section{Eulerian Poisson brackets}\label{app:Eulerian PBs}

We now calculate the \textit{Eulerian} PBs for 3D systems by substituting the \textit{canonical} conjugate pairs $(\bm{R}^\alpha,\bm{P}^\alpha)$ and $(\bm{\theta}^\alpha,\tilde{\bm{\ell}}^\alpha)$, derived in Appendix~\ref{app:canonical conjugate pair}, into the PB definition of Eq.~\eqref{eq:PB expression}. We focus on the less familiar PBs that are specific to micropolar elasticity, namely those involving the displacement field $\bm{u}$ and the internal rotation field $\bm{\theta}$. The remaining PBs have already been derived in the literature~\cite{stark2003,stenull2004,stark2005,markovich2021,markovich2024}. At the end of this section, we list these Eulerian PBs, their reduced forms for 2D systems, and their comparison with the corresponding Lagrangian PBs.

We first calculate the PB between the angular momentum density $\bm{\ell}(\bm{R})$ and the internal rotation $\bm{\theta}(\bm{R}')$. Since this bracket concerns only rotational degrees of freedom, it involves only the microscopic canonical conjugate pair $(\bm{\theta}^\alpha,\tilde{\bm{\ell}}^\alpha)$:
\begin{align}
\{\ell_i (\bm{R}) , \theta_j(\bm{R'})\} 
&= \sum_{\alpha} 
\frac{\partial \ell_i(\bm{R})}{\partial\tilde\ell^\alpha_k}
\frac{\partial \theta_j(\bm{R}')}{\partial\theta^\alpha_k}
\notag
\\
&=
\sum_\alpha M_{ij}^\alpha\frac{1}{n(\bm{R}')}\delta(\bm{R}-\bm{R}^\alpha)\delta(\bm{R}'-\bm{R}^\alpha)%
\notag\\
&= M_{ij}(\bm{R}') \delta(\bm{R}-\bm{R}')
\label{appeq:PB ell theta}
\, .
\end{align}
In the second line of Eq.~\eqref{appeq:PB ell theta}, we apply the chain rule and then use the delta-function identity $\delta(\bm{R}-\bm{R}^\alpha)\delta(\bm{R}'-\bm{R}^\alpha)=\delta(\bm{R}-\bm{R}')\delta(\bm{R}'-\bm{R}^\alpha)$ to obtain the final expression. We define the particle-averaged field $\bm{M}(\bm{R})\equiv\sum_\alpha \bm{M}^\alpha\delta(\bm{R}-\bm{R}^\alpha)/n(\bm{R})$, in the same way as the displacement $\bm{u}(\bm{R})$ and the internal rotation $\bm{\theta}(\bm{R})$ in Eq.~\eqref{eq:particle-averaged fields}. Here, $\bm{M}^\alpha\equiv\bm{\mathcal{M}}(\bm{\psi}=\bm{\theta}^\alpha)$ is the matrix microscopically relating the usual and canonical angular momenta in Eq.~\eqref{appeq:M}, evaluated at the internal rotation of particle $\alpha$.

In the \textit{Eulerian} framework, fields are defined using the \textit{real} particle positions through the delta function $\delta(\bm{R}-\bm{R}^\alpha)$. Therefore, the CM momentum density $\bm{g}^c(\bm{R})\equiv\sum_\alpha \bm{P}^\alpha \delta(\bm{R}-\bm{R}^\alpha)$ has a non-vanishing PB with the internal rotation field $\bm{\theta}(\bm{R}')$:
\begin{align}
\{ g_i^c (\bm{R}) , \theta_j(\bm{R}')  \} 
&=
\sum_{\alpha} 
\frac{\partial g_i^c(\bm{R})}{\partial P_k^\alpha} 
\frac{\partial \theta_j (\bm{R}')}{\partial R_k^\alpha} 
\notag\\
&=
\frac{1}{n^2(\bm{R}')}
\sum_\alpha
\delta(\bm{R}-\bm{R}^\alpha)
\bigg\{
\bigg[
\sum_\beta
\theta^\beta_j\delta(\bm{R}'-\bm{R}^\beta)
-n(\bm{R}')\theta^\alpha_j
\bigg]\nabla_i'\delta(\bm{R}'-\bm{R}^\alpha)
\bigg\}
\notag\\
&=
\frac{1}{n(\bm{R}')}
\big\lbrace
-\nabla'_i\big[
\theta_j(\bm{R}')n(\bm{R}')\delta(\bm{R}-\bm{R}')
\big]
+\theta_j(\bm{R}')\nabla_i'\big[
n(\bm{R}')\delta(\bm{R}-\bm{R}')
\big]
\big\rbrace
\notag\\
&=
-\big[\nabla'_i \theta_j(\bm{R}') \big]\delta(\bm{R}-\bm{R}')
\label{appeq:PB gc theta}
\, .
\end{align}
Here, we have used $\partial \delta(\bm{R}'-\bm{R}^\alpha)/\partial R_i^\alpha=-\nabla_i'\delta(\bm{R}'-\bm{R}^\alpha)$. 

As in the calculation of $\{g_i^c(\bm{R}),\theta_j(\bm{R}')\}$, the PB between the CM momentum density $\bm{g}^c$ and the displacement field $\bm{u}(\bm{R}')$ contains an additional gradient term $\nabla'_i u_j(\bm{R'})$ arising from $\partial \delta(\bm R'-\bm R^\alpha)/\partial \bm R^\alpha$, in addition to  the trivial Kronecker-delta contribution from $\partial u^\alpha_j/\partial R^\alpha_i = \partial (R^\alpha_j - r^\alpha_j) /\partial R^\alpha_i = \delta_{ij}$:
\begin{align}
\{  g_i^c(\bm{R}) , u_j(\bm{R}')  \} 
&
= \sum_{\alpha} 
\frac{\partial g_i^c(\bm{R})}{\partial P_k^\alpha} 
\frac{\partial u_j(\bm{R}')}{\partial R_k^\alpha}
\notag
\\
&
= 
\frac{1}{n(\bm{R}')}
\sum_\alpha
\delta(\bm{R}-\bm{R}^\alpha)
\big\lbrace
\delta_{ij}\delta(\bm{R}'-\bm{R}^\alpha)
+
\big[u_j(\bm{R}')-u^\alpha_j\big]
\nabla'_i 
\delta(\bm{R}'-\bm{R}^\alpha)
\big\rbrace
\notag\\
&=
\big[
\delta_{ij}
-\nabla'_i u_j(\bm{R}')
\big]\delta(\bm{R}-\bm{R}')
\label{appeq:PB gc u}
\, .
\end{align}
The gradient terms generated by differentiating the Eulerian delta function in Eqs.~\eqref{appeq:PB gc theta} and \eqref{appeq:PB gc u} are characteristic Eulerian contributions~\cite{forster1974,lubensky1985,kamien2000,stark2003,stark2005,hernandez2021,triguero-platero2023,markovich2021,markovich2024}, which are absent from the corresponding \textit{Lagrangian} PBs and therefore represent one of the main differences between the two PB formalisms.

Although not related to the PBs of $\bm\theta(\bm{R})$ and $\bm{u}(\bm{R})$,  
the PB between the angular momentum $\bm\ell$ itself is worth calculating using the conjugate pair $(\bm\theta,\tilde{\bm{\ell}})$, which reproduces the known result in the  literature~\cite{markovich2021,markovich2024,lubensky2005}:  
\begin{align}
\notag\{\ell_i (\bm{R}) , \ell_j(\bm{R}')\} 
&= \sum_{\alpha} 
\frac{\partial \ell_i(\bm{R})}{\partial\tilde\ell^\alpha_k}
\frac{\partial \ell_j(\bm{R}')}{\partial\theta^\alpha_k} %
-
\frac{\partial \ell_i(\bm{R})}{\partial\theta^\alpha_k}
\frac{\partial \ell_j(\bm{R}')}{\partial\tilde\ell^\alpha_k}
\notag\\ %
&= \sum_\alpha
\frac{\partial \ell_i(\bm{R})}{\partial\ell^\alpha_m}
\frac{\partial \ell^\alpha_m(\bm{R})}{\partial\tilde\ell^\alpha_k}
\frac{\partial \ell_j(\bm{R}')}{\partial\theta^\alpha_k}
-\frac{\partial \ell_i(\bm{R}')}{\partial\ell^\alpha_m}
\frac{\partial \ell^\alpha_m(\bm{R}')}{\partial\tilde\ell^\alpha_k}
\frac{\partial \ell_j(\bm{R})}{\partial\theta^\alpha_k}
\notag\\%
&
=\sum_{\alpha}M_{ik}^\alpha\delta(\bm{R}-\bm{R}^\alpha)%
\left[\tilde\ell_p^\alpha \frac{\partial M^\alpha_{jp}}{\partial \theta_k^\alpha} \delta(\bm{R}'-\bm{R}^\alpha)\right]%
-M_{jk}^\alpha\delta(\bm{R}'-\bm{R}^\alpha)%
\left[\tilde\ell_p^\alpha \frac{\partial M^\alpha_{ip}}{\partial \theta_k^\alpha} \delta(\bm{R}-\bm{R}^\alpha)\right]\notag\\%
&
=
\sum_\alpha
\tilde\ell_p^\alpha 
\bigg(   
M_{ik}^\alpha\frac{\partial M_{jp}^\alpha}{\partial \theta_k^\alpha}
-
M_{jk}^\alpha\frac{\partial M_{ip}^\alpha}{\partial \theta_k^\alpha}
\bigg)
\delta(\bm{R}-\bm{R}')\delta(\bm{R}'-\bm{R}^\alpha) 
\notag\\
&= 
-\sum_\alpha  
\varepsilon_{ijm}
M_{mp}^\alpha\tilde\ell_p^\alpha
\delta(\bm{R}-\bm{R}')\delta(\bm{R}'-\bm{R}^\alpha) 
\notag\\
&=
-\varepsilon_{ijk}\ell_k(\bm{R}')\delta(\bm{R}-\bm{R}')
\, ,
\label{appeq:ell ell PB}
\end{align}
where we have used 
the definition $M^\alpha_{mp}\tilde\ell^\alpha_p =\ell^\alpha_m$. To calculate the term in parentheses in the fourth line we use Eq.~\eqref{appeq:M} to write, 
\begin{align}
M^\alpha_{ik}\frac{\partial M^\alpha_{jp}}{\partial \psi_k} 
&=
\frac{f}{2}\varepsilon_{ipj}
+
\big(
fg+\frac{1}{4}
\big)
\delta_{ip}\psi_j
+
\big(
fg-\frac{1}{4}
\big)
\delta_{ij}\psi_p
+
\frac{f'}{\psi}\delta_{pj}\psi_i
\notag
\\
&+
\big(
\frac{g'}{\psi}
+
2g^2
\big)
\psi_i\psi_p\psi_j
+
\frac{g}{2}
\big(
\psi_i\varepsilon_{pjq}\psi_q
+
\varepsilon_{piq}\psi_q\psi_j
+
\varepsilon_{jiq}\psi_q\psi_p
\big)
\label{appeq:M derivative}
\, .
\end{align}
Here ${\bm \psi}={\bm \theta}^\alpha$, $f\equiv(\psi/2)\cot(\psi/2)$, $g\equiv(1-f)/\psi^2$, $f'\equiv d f/d \psi$ and $g' \equiv d g/d\psi$. Equation~\eqref{appeq:M} can further be expressed as $M^\alpha_{ij}=f\delta_{ij}
+g\psi_i\psi_j-(\varepsilon_{ijk}\psi_k/2)$. Exchanging the indices $i\leftrightarrow j$ in Eq.~\eqref{appeq:M derivative} we find
\begin{align}
M^\alpha_{ik}
\frac{\partial M^\alpha_{jp}}{\partial \psi_k}
-
M^\alpha_{jk}
\frac{\partial M^\alpha_{ip}}{\partial \psi_k}
&=
-\varepsilon_{ijm}
\big(
f\delta_{pm}
+
g\psi_p\psi_m
+
\frac{1}{2}\varepsilon_{pmq}\psi_q
\big)
=
-\varepsilon_{ijm} M^\alpha_{mp}
\label{appeq:M derivative diff}
\, , 
\end{align}
where all terms symmetric in $i\leftrightarrow j$ in Eq.~\eqref{appeq:M derivative} cancel off.
%


Finally, we collect the non-zero fundamental PBs of the Eulerian framework:
\begin{align}
\lbrace \ell_i(\bm{R}),\theta_j(\bm{R}') \rbrace
&= 
M_{ij}(\bm{R}')\delta(\bm{R}-\bm{R}'), 
\label{appeq:ell theta in list}
\\
\lbrace \ell_i(\bm{R}),\ell_j(\bm{R}') \rbrace
&= 
-\varepsilon_{ijk} \ell_k(\bm{R}')\delta(\bm{R}-\bm{R}') 
,
\\
\lbrace g_i^c(\bm{R}),\ell_j(\bm{R}') \rbrace    
&= 
-\nabla_i'\left[\ell_j(\bm{R}')\delta(\bm{R}-\bm{R}')\right] 
,
\\
\lbrace g_i^c(\bm{R}),\theta_j(\bm{R}') \rbrace    
&=
-\big[\nabla_i' \theta_j (\bm{R}') \big]\delta(\bm{R}-\bm{R}')
,
\\
\lbrace g_i^c(\bm{R}), \rho(\bm{R}') \rbrace
&=
-\nabla_i'\left[\rho(\bm{R}')\delta(\bm{R}-\bm{R}')\right]
,
\\
\lbrace g_i^c(\bm{R}), u_j(\bm{R}') \rbrace    
&=    
\big[\delta_{ij}
-\nabla'_i u_j(\bm{R}')
\big] \delta(\bm{R}-\bm{R}') 
,
\\
\lbrace g_i^c(\bm{R}), g_j^c(\bm{R}') \rbrace 
&=
g_i^c(\bm{R}')\nabla_j\delta(\bm{R}-\bm{R}') 
-g_j^c(\bm{R})\nabla'_i\delta(\bm{R}'-\bm{R}) 
,
\\
\lbrace{g_i^c(\bm{R}),\tau_j(\bm{R}') \rbrace}
&=- \nabla_i'\left[
\tau_j(\bm{R}')\delta(\bm{R}-\bm{R}')
\right]
\label{appeq:gc tau in list}
.
\end{align}
For comparison, we also list the only three non-vanishing \textit{Lagrangian} PBs:
\begin{align}
\lbrace \ell_i^\circ(\bm{r}),\theta_j(\bm{r}') \rbrace
&= 
M_{ij}(\bm{r}')\delta(\bm{r}-\bm{r}'), 
\\
\lbrace \ell_i^\circ(\bm{r}),\ell_j^\circ(\bm{r}') \rbrace
&= 
-\varepsilon_{ijk} \ell_k^\circ(\bm{r}')\delta(\bm{r}-\bm{r}') 
,
\\
\lbrace g_i^{\circ}(\bm{r}), u_j(\bm{r}') \rbrace    
&=    
\delta_{ij}
\delta(\bm{r}-\bm{r}') 
.
\end{align} 
In the 2D case, the internal rotation and angular momentum reduce to scalars $(\bm{\theta}^\alpha,\bm{\ell}^\alpha)\rightarrow(\theta^\alpha,\ell^\alpha)$ and $M_{ij}= M^{-1}_{ij} = \delta_{ij}$ because there is only one rotational direction. As a result, the PBs involving the internal rotation and angular momentum density become $\lbrace \ell(\bm{R}),\theta(\bm{R}')\rbrace=\delta(\bm{R}-\bm{R}')$ and $\lbrace \ell(\bm{R}),\ell(\bm{R}')\rbrace=0$ in the Eulerian framework. Similarly, in the Lagrangian framework, they reduce to $\lbrace \ell^\circ(\bm{r}),\theta(\bm{r}')\rbrace=\delta(\bm{r}-\bm{r}')$ and $\lbrace \ell^\circ(\bm{r}),\ell^\circ(\bm{r}')\rbrace=0$~\cite{lee2026a}.

\section{Eulerian Dynamics for 2D Chiral Active Solids}\label{app:2D dynamic eqs}

Here we derive the \textit{Eulerian} field dynamics of the 2D chiral active solid studied in the main text. We use the PB formalism of Eq.~\eqref{eq:PB integral}, the Eulerian Hamiltonian of Eq.~\eqref{eq:H Eulerian} with the potentials specified in Eqs.~\eqref{eq:Eulerian Vtot} and \eqref{eq:VE}, and the 2D Eulerian PBs listed in Eqs.~\eqref{appeq:ell theta in list}--\eqref{appeq:gc tau in list}. 

We begin with the fields whose Eulerian dynamics are purely \textit{kinematic}: the displacement $\bm{u}$, the internal rotation $\theta$, the density $\rho$, and the torque density $\tau$. Their non-vanishing PBs involve only the CM momentum density $\bm{g}^c$ or the angular momentum density $\ell$. Therefore, their dynamics require only the Hamiltonian derivatives associated with these kinetic fields, namely $\delta H/\delta g_j^c=v_j^c$ and $\delta H/\delta \ell=\Omega$. Substituting these derivatives into Eq.~\eqref{eq:PB integral} yields $\dot{u}_i+v_j^c\nabla_j u_i=v_i^c$, $\dot{\theta}+v_j^c\nabla_j\theta=\Omega$, $\dot{\rho}+\nabla_j(\rho v_j^c)=0$, and $\dot{\tau}+\nabla_j(\tau v_j^c)=0$. These dynamic equations involve only the particle velocity $v_i^c$, the angular velocity $\Omega$, and  streaming terms, with no contributions from the elastic potential. 

The elastic potential contributes only to the dynamics of the CM momentum density and angular momentum density. Inserting the PBs involving $\bm{g}^c$ and $\ell$ into the PB integral in Eq.~\eqref{eq:PB integral} gives the general dynamical forms in Eqs.~\eqref{eq:gc dynamics} and \eqref{eq:Eulerian ell dynamics}. In deriving these equations, we have also used $\delta H/\delta \rho=-\frac{1}{2}[(\bm{v}^c)^2+\ell\Omega/\rho]$, which has a contribution from the rotational kinetic energy because the particles are assumed identical such that the moment-of-inertia density $I\propto\rho$.

While the calculation of $\delta H/\delta\theta$ for $\dot{\ell}$ in Eq.~\eqref{eq:Eulerian ell dynamics} is straightforward, deriving the CM momentum dynamics $\dot{\bm{g}}^c$ (Eq.~\eqref{eq:gc dynamics}) requires a few additional steps. We start from writing the explicit form of the CM momentum dynamics:
\begin{align}
&\dot{g}_i^c + \nabla_j(v_j^c g_i^c)
=
2\kappa_c 
\phi\nabla_i\big[
\phi + \frac{1}{2}\nabla\times\bm{u}
\big]
+
\big[\nabla\times(\kappa_c\phi)\big]_i
-\kappa_c(\nabla\times\phi)_j\nabla_i u_j
+ E_{ijkl}\nabla_j\nabla_l u_k  
\notag
\\
&
+\nabla_j
\bigg\lbrace
-(\tilde\lambda+\tilde\mu)\phi^2\delta_{ij}
+\phi\bigg[
-(\tilde\lambda+\frac{\kappa_c}{2})\varepsilon_{ij}\delta_{kl}
+\tilde\mu \varepsilon_{kl}\delta_{ij}
+ (\kappa_c-\tilde\mu)(\varepsilon_{il}\delta_{jk}+\varepsilon_{kj}\delta_{il})
\bigg]
\nabla_l u_k
\bigg\rbrace
\, .
\label{appeq:step to Eulerian gc}
\end{align}
Note that in calculating $\dot{\bm g}^c$ (Eq.~\eqref{appeq:step to Eulerian gc}), the contributions from the active potential $-\tau\theta$ cancel off in calculating $(\delta H/\delta\theta) {+} \tau$. Therefore, $\dot{\bm g}^c$ is purely elastic and Eq.~\eqref{appeq:step to Eulerian gc} can be reorganized into a proper stress-divergence form,
$\dot{g}_i^c+\nabla_j(v_j^c g_i^c)=\nabla_j\sigma_{ij}$. 
To achieve this, we use the identity
\begin{equation}\label{appeq:grad curl to divergence}
\phi\nabla_j(\nabla\times\bm{u})-(\nabla\times\phi)_j\nabla_i u_j = \nabla_j[\varepsilon_{jk}\delta_{il}\phi\nabla_l u_k] 
\, ,
\end{equation}
where 
\begin{align}
\varepsilon_{jk}\delta_{il} 
= 
\frac{1}{4}(
\varepsilon_{jk}\delta_{il}
+\varepsilon_{jl}\delta_{ik}
+\varepsilon_{ik}\delta_{jl}
+\varepsilon_{il}\delta_{jk})
-\frac{1}{2}\varepsilon_{ij}\delta_{kl}
-\frac{1}{2}\varepsilon_{kl}\delta_{ij}
\label{appeq:symmetry decomp}
\, .    
\end{align}
Equation~\eqref{appeq:symmetry decomp} is also used for the last two terms in the square brackets of Eq.~\eqref{appeq:step to Eulerian gc} (also with the interchange  $(i,l) \leftrightarrow (j,k)$).
%
To derive Eq.~\eqref{appeq:symmetry decomp}, we first separate the tensor $\varepsilon_{jk}\delta_{il}$ into its symmetric and antisymmetric parts with respect to the interchange $k\leftrightarrow l$, then use the same procedure for $i\leftrightarrow j$, and then apply  $\varepsilon_{jk}\delta_{il}-\varepsilon_{jl}\delta_{ik}
=\varepsilon_{jm}(\delta_{mk}\delta_{il}-\delta_{ml}\delta_{ik})
=(\varepsilon_{jm}\varepsilon_{mi})\varepsilon_{kl}
=-\varepsilon_{kl}\delta_{ij}$ and similarly $\varepsilon_{jk}\delta_{il}-\varepsilon_{ik}\delta_{jl}=-\varepsilon_{ij}\delta_{kl}$.
%
%

Taken together, we recover Eq.~\eqref{eq:gc dynamics}, where the combination $\varepsilon_{il}\delta_{jk}+\varepsilon_{jk}\delta_{il}$ in Eq.~\eqref{appeq:step to Eulerian gc} gives only non-odd terms. 
The remaining odd term in Eq.~\eqref{eq:gc dynamics} originates solely from $\varepsilon_{jk}\delta_{il}$ in Eq.~\eqref{appeq:grad curl to divergence}, which is a result of the combined spatial-derivative terms $[\nabla_j u_i(\bm{R})]\delta(\bm{R}-\bm{R}')$ and $-\big[\nabla_i' \theta (\bm{R}') \big]\delta(\bm{R}-\bm{R}')$ in the Eulerian PBs.
In deriving Eq.~\eqref{eq:gc dynamics} we have also used the identity  
~\cite{scheibner2020,fruchart2023, markovich2025}: 
\begin{equation}
(\varepsilon_{jk}\delta_{il}
+\varepsilon_{jl}\delta_{ik}
+\varepsilon_{ik}\delta_{jl}
+\varepsilon_{il}\delta_{jk})\nabla_l u_k 
=
2 (\varepsilon_{jl}\delta_{ik}+\varepsilon_{ik}\delta_{jl})\nabla_l u_k    
\, .
\end{equation}

\bibliographystyle{iopart-num}
\bibliography{references}

\end{document}